\documentclass{ws-jai}
\usepackage[flushleft]{threeparttable}
\pdfoutput=1

\begin{document}

\catchline{}{}{}{}{} 

\markboth{R. K. Banyal \& A. Reiners}{A dual cavity Fabry-Perot device for high precision Doppler measurements in astronomy}

\title{A dual cavity Fabry-Perot device for high precision Doppler measurements in astronomy}

\author{R. K. Banyal$^\dagger$ and A. Reiners$^\ddagger$}

\address{
$^\dagger$Indian Institute of Astrophysics, Bangalore, 560034 INDIA, banyal@iiap.res.in\\
$^\ddagger$Institut f\"{u}r Astrophysik, Friedrich-Hund-Platz 1, 37077 G\"{o}ttingen, Germany\\
}

\maketitle

\corres{$^\dagger$Corresponding author.}

\begin{history}
\received{(to be inserted by publisher)};
\revised{(to be inserted by publisher)};
\accepted{(to be inserted by publisher)};
\end{history}

\begin{abstract}
We propose a dual cavity Fabry-Perot interferometer as a wavelength calibrator and a stability tracking device for astronomical spectrograph. The FPI consists of two adjoining cavities engraved on a low expansion monoblock spacer. A low-finesse astro-cavity is intended for generating a uniform grid of reference lines to calibrate the spectrograph and a high-finesse lock-cavity is meant for tracking the stability of the reference lines using optical frequency standards. The differential length changes in two cavities due to temperature  and vibration perturbations are quantitatively analyzed using finite element method. An optimized mounting geometry with fractional length changes  $\Delta L/L \approx 1.5\times 10^{-12}$ is suggested. We also identify conditions necessary to suppress relative length variations between two cavities  well below 10$^{-10}$~m, thus facilitating accurate dimension tracking and generation of stable reference spectra for Doppler measurement at 10 cms$^{-1}$ level.
\end{abstract}


\section{Introduction}
High precision Doppler spectroscopy is an important tool to investigate many astrophysical processes in the Universe. Measurement of
tiny Doppler shifts from radial velocity (RV) can provide insight about the stellar interior; reveal the existence of unseen planets around other stars; help us probe the variability of fundamental constants of nature and provide useful information of binary and pulsating stars \cite{san08}. In particular, the radial velocity technique has been immensely successful in detecting hundreds of planets with different mass range and orbital periods \cite{mar11}. The main challenge and focus of contemporary research is to find Earth-analogs within the habitable zone of star where life sustaining conditions could prevail. Recently, the Doppler data from ESO's two precision RV instruments, namely HARPS and UVES, was decisive in unveiling the presence of 1.3 Earth-mass rocky planet orbiting Proxima Centauri - the nearest known star to the Sun \cite{gui16}.

Typically, Doppler precision of  $\simeq 10$ cms$^{-1}$ or better is required to address many compelling science cases including the  detection of terrestrial-size planets around sun-like stars \cite{uza11,gri09}. Leading RV observing campaigns are carried out with state-of-the-art echelle spectrograph, carefully designed and built to operate under tightly regulated thermovac conditions \cite{pep14,fis16}. A high resolution spectrum of star in visible/NIR band is recorded on a 2D array of CCD pixels. To extract the RV signal from time series data, spectrograph is wavelength calibrated with a laboratory source of well known emission or absorption lines. In calibration process several lines of known wavelengths are identified and a quadratic or high order polynomial fit is used to invert the pixel coordinates into accurate wavelength units.

Two commonly used wavelength reference in astronomy are thorium-Argon (Th-Ar) emission lamp and Iodine absorption cell \cite{fis14}. Despite averaging the shift over thousands of spectral lines, the best RV precision obtained from these sources  is limited to $\sim1$~ms$^{-1}$. The RV uncertainty of traditional sources arises from wavelength inaccuracies of individual lines, spectral contamination, aging of the lamp, limited spectral coverage, line blending effects,  uneven distribution and varying intensity of lines across spectrograph band \cite{mur07}.

An ideal calibrator choice for enabling sub-ms$^{-1}$ RV precision in next generation spectrographs is  a broadband laser frequency comb (LFC). The LFC consists of thousands of equispaced lines produced from a femtosecond mode-locked laser \cite{mur07,ste08}. The entire comb spectrum is stabilized by anchoring to atomic clock with well established radio-frequency locking technique. Since past decade different versions of LFCs have been tested and validated \cite{qui10,gab12,pro14}. New developments are still awaited to overcome a couple of technical and operational difficulties. It might take a while for LFC to become a simple and affordable turn-key instrument for routine wavelength calibration at most observatories.

A simple and cost effective approach to produce reference spectra is to use cavity resonance lines of a FP etalon \cite{wil12}. A plane parallel FP cavity when illuminated with a broadband white light source acts as a periodic wavelength filter. The transmitted light consists of evenly distributed lines that result from multiple beam interference inside the cavity \cite{vau89}. The cavity lines are separated by free spectra range $\textrm{FSR}=c/2L$, where $c$ is the light speed and $L$ is the cavity length.  Many variants of FPs (e.g. solid, air spaced and fiber based) have been successfully built and  tested for astronomy calibration \cite{sch12,hav14,fur14,bau15}.

The positional accuracy of FP lines critically depends on stability of the cavity itself. Any noise affecting the cavity length would randomly displace the resonant lines thus directly impairing the achievable RV precision. The impact of temperature and pressure fluctuations on the cavity length is generally eliminated by mounting the FP inside a thermally controlled evacuated chamber which is capable of achieving $\sim$~1 ms~$^{-1}$ short-term stability \cite{sch12}.

A passively stabilized FP apparatus producing reference spectra has no built--in mechanism to track systematic offsets or drifts in cavity length occurring over time scales of weeks, months or even years. Regardless of good passive isolation, cavity drift could still occur from temperature residuals, material aging or changes in optical properties of the coatings \cite{rei98,fox09,den14}. The radial velocity $v$ and the fractional length and frequency stability of the cavity can be expressed as $v/c=\Delta L/L=\Delta\nu/\nu$. For instance, a 1~nm change in a 1~cm long cavity corresponds to $\sim$55~MHz frequency shift of cavity lines in visible ($\lambda=550$ nm) region, which amounts to 30~ms$^{-1}$ RV uncertainty. To reach a RV precision $\leq10$ cms$^{-1}$, the cavity length should be stable to $\sim10^{-10}$ level or better.

For  long-term usability (several months to years) of astro-FP, research efforts are now being directed at controlling and tracking the time dependent cavity drift with advance techniques \cite{ren14,sch14,huk15,stu16}. Measurement of tiny dimensional changes in cavity length calls for advanced metrology  based on laser frequency standards. A drift tracking or active length-control  requires cavity-locking to a frequency stabilized diode laser or vice-versa \cite{ren14}. In cavity-lock setup, the frequency stability of  the laser is expected to be transferred to the stability of the FP cavity lines. The main task of frequency or phase locking mechanism is to keep cavity transmission point in resonance with frequency of the laser \cite{dre83}.

A short-length, low-finesse($\mathcal{F}\approx10$) cavity, while useful for generating  broadband reference lines for astronomy spectrographs, is not ideal for laser metrology. Normally, a robust laser-lock requires a sharp discriminant signal which depends on the laser power and the cavity linewidth \cite{bla01}. The linewidth $\Delta\nu=\textrm{FSR}/\mathcal{F}$ of the cavity should be sufficiently narrow to produce a steeper error signal.  The inverse dependence of $L$ and $\mathcal{F}$ on the linewidth  $\Delta\nu$ implies a clear advantage of longer  and high finesse cavities.  Another practical difficultly of utilizing astro-FP in laser applications is the cavity geometry. A confocal cavity design is more appropriate for locking purpose because it greatly simplifies beam alignment and mode-matching conditions for single frequency laser beam. On the contrary, a collimated white light beam is easily coupled into a short-length planar cavity producing reference spectra \cite{huk15}. These considerations show incompatible requirements of a cavity for laser metrology and astronomy calibration. Clearly, a single FP cavity is inadequate for serving dual purpose of `\emph{wavelength calibration}' on one hand and `\emph{drift tracking}' on other.

We propose to circumvent this problem  by constructing two adjoining cavities of equal length but of different finesse on a single block of ultra low expansion (ULE) glass. The low finesse \textit{astro-cavity} would produce the calibration spectra while the high finesse \textit{lock-cavity}  would facilitate frequency locking to implement stability tracking mechanism. The length of two cavities is defined by the length of monolithic spacer block. Here, the implicit assumption is that external perturbations, if any, act uniformly on the entire FP block, thus causing equal and identical length changes in both cavities. In fact, drift measurements would only be accurate so long the relative length variation between two cavities is well below $10^{-10}$. Under this premise alone the \emph{lock-cavity} would function as a reliable proxy for measuring the stability of the \emph{astro-cavity}. Departure from this scenario would result in intractable measurement errors.
\begin{figure}[htbp]
\centerline{\includegraphics[width=0.65\columnwidth]{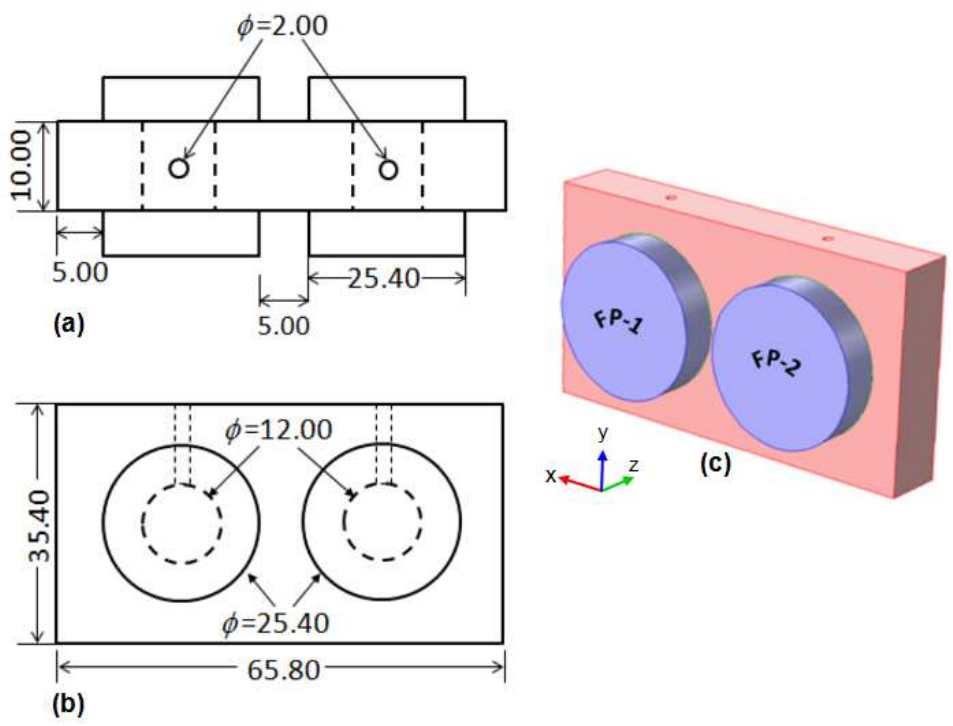}}
\caption{The geometry of a dual channel FP resonator. All relevant dimensions are shown in mm. (a) The top view showing the 10 mm thick cavity spacer with optically contacted mirror pairs on either side. Two small holes of $\phi=2$~mm  on top are provided for the venting purpose. (b) The front view showing the spacer dimensions along with two cavities of diameter $\phi=12$~mm each and standard HR mirrors with $\phi=25.4$~mm each. (c) The isometric view. The cavity axis is orientated along the $z$-direction and the lab vertical is along $y$-direction.}
\end{figure}

The main objective of this paper is to numerically examine the validity of our approach. We identify the temperature and vibration induced noise conditions that impart unequal length changes in two adjoining cavities constructed on a monoblock spacer. These studies are important to understand the level of performance one can realistically expect to achieve in the proposed device. We use finite element method to numerically simulate the impact of small thermal gradients and orientation dependent vibration noise that could deforms two cavities in dissimilar manner.

From here on, the paper is organized as follows. The dual FP design and laser-lock setup for cavity drift tracking is discussed in Section 2. In Section 3, the mathematical formulism and FEA model for cavity deformation is described. The vibration analysis of horizontally mounted cavity is presented in Section 4. A proposed vertical mounting configuration  and numerical optimization procedure to reduce the vibration sensitivity of the FP device is given in Section 5. Thermo-elastic model is used to calculate the temperature sensitivity of the FP device. Effect of temperature gradient and cavity length offset is presented in Section 6. Numerical accuracy of FEA results is discussed in Section 7 and finally, the summary of the work is given in Section 8.

\section{Dual Cavity FP Design}
Fabry-Perot cavities are used in diverse range of applications in many areas of fundamental and applied research \cite{jun03}. Ultrastable cavities are also key to gravitational wave detection, testing of Lorentz invariance and experiments in future space missions \cite{abb16,her09,arg12}. For wavelength calibration, the FP cavity spacing and mirror reflectivity $R$ are selected based on  spectrograph requirements. The cavity spacing $L$ is chosen so that reference lines are easily resolved by spectrograph in the detector plane.  Generally, a short cavity (e.g., $L=1$ cm, $\textrm{FSR}=c/2L=15$ GHz) is required to generate optimally spaced comb lines to calibrate a high resolution ($\lambda/\Delta\lambda\sim100,000$) astronomy spectrograph. The reflectivity ($R\sim70\%$) of broadband coated mirrors yields a low finesse $\mathcal{F}\sim10$. The resulting cavity linewidth $\Delta \nu=FSR/\mathcal{F}$ is such that it is optimally sampled by the detector pixels. To produce reference lines that are 15 GHz apart would  require a planar cavity of length 1~cm.

\subsection{Geometry}
The main component of the proposed device is a two-channel FP etalon. It consists of two cylindrical cavities of diameter 12~mm each, bored adjacent to each other on a rectangular block ($10\times35.4\times65.8$ mm$^{3}$) of ultra low expansion (ULE) glass. The cavity mirrors are made of fused silica (FS) and optically contacted  on either side of the spacer block. We assume the optical contacts to be sufficiently strong and free from any imperfections. This is a reasonable assumption as the modern techniques of joining optical elements have become more robust \cite{hai07}. A CAD geometry of the device is shown in Fig.~1. The low-finesse \textit{astro-cavity} (FP1) is to provide astronomy calibration lines and high-finesse \textit{lock-cavity} (FP2) is to provide a light channel for robust laser-lock. The cavity length $L_{1}=L_{2}$=1~cm is chosen to produce 15~GHz apart reference lines to calibrate a typical high resolution (100,000) Echelle spectrographs.  To maintain sufficient rigidity of the entire structure, a small 5~mm gap is provided on all sides between mirrors and spacer edges. Cavity mirrors are assumed to be flat. Replacing flat mirrors with curved mirrors will not alter our analysis in any major way, except a very small change in curvature due to differential thermal expansion.

\begin{figure}[htbp]
\centerline{\includegraphics[width=0.7\columnwidth]{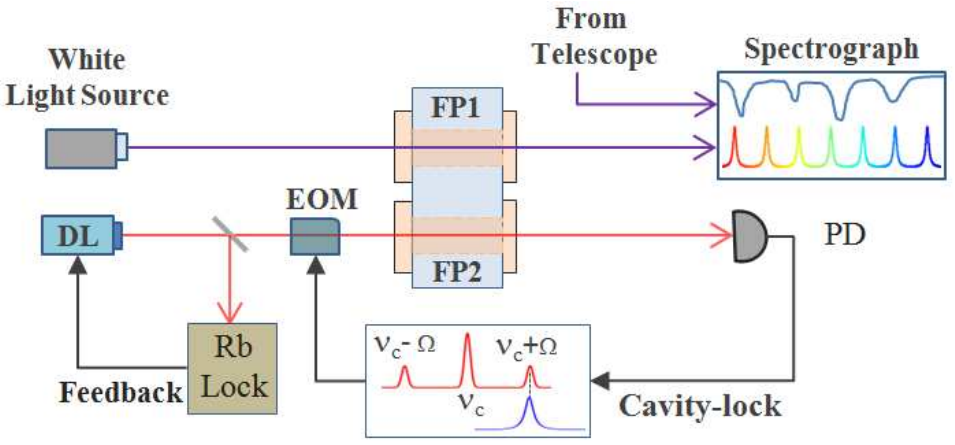}}
\caption{Schematic of dual FP cavity-lock set up. A collimated white light beam from halogen lamp or laser driven white light source, feeds FP1 to generates wideband comparison spectra for astronomy spectrograph. The high-finesse FP2 facilitates  cavity drift tracking. The frequency of diode laser (DL) is stabilized by externally locking it to 780~nm Rb87 D2 line using standard saturation absorption technique. The cavity resonance is probed by one of the laser sidebands $\nu_c\pm \Omega$ generated by electro-optic modulator (EOM). The RF electronics following the photodiode (PD) provides necessary feedback to keep one of the sideband frequencies, e.g. $\nu_c+ \Omega $ (shown in red), locked to the peak of cavity transmission line (shown in blue). For clarity, other ancillary components are not shown.}
\end{figure}

\subsection{Laser-lock and Drift Tracking Mechanism}
The conceptual principle of dual FP cavity-lock is illustrated in Fig. 2. The cavity drift can be tracked in several ways, but often requires two lasers \cite{rei98}. However, the same goal can be accomplished using a simpler approach of \emph{sideband locking} which needs only one laser \cite{tho08,liv09}. In this method the frequency $\nu_{c}$ of the laser is pre-stabilized to Rb-cell using standard saturation absorption \cite{wie91}. A tunable electro-optic modulator (EOM) creates  radio frequency (RF) sidebands $\nu_c\pm \Omega$ around the central frequency $\nu_{c}$ of the laser. One of the sidebands is then locked to the nearest resonance peak of FP2 cavity using modified Pound-Drever-Hall (PDH) method. The RF frequency $\Omega$ of the oscillator driving EOM is  servo-locked to cavity.  This technique preserves the intrinsic stability of main laser beam $\nu_{c}$ that is locked to rubidium while exploiting the sideband tunability to track the cavity drift.

To estimate the drift, the frequency of the radio oscillator has to be readout independently by a frequency counter or RF spectrum analyzer. A fractional length change ($\Delta L/L=\Delta \nu/\nu=v/c$) about $10^{-10}$ of the cavity block corresponds to RV error of 10~cms$^{-1}$ or a RF frequency shift of few Hz in an EOM operating at 1~GHz.  Measuring a GHz RF signal with an accuracy of few Hz (amounting to few ppb frequency resolution) is a challenging task. However, the desired precision can be reached with many cycles of repeated measurements.  Alternatively, optical-beat measurement with two lasers yields much better sensitivity when the cavity is used in reflection mode, i.e. PDH locking \cite{ren14,dav07}. For example, a fractional change $10^{-10}$ of the cavity length would induce a readily measurable frequency excursion of $\sim40$~kHz in the optical-beat signal obtained with PDH setup using 780 nm laser. The exact implementation details will be worked out in a separate study.

\section{Cavity Deformation}
The resonance stability of  FP cavity is  determined by the stability of optical length between each mirror pairs. Different noise sources e.g., seismic and acoustic vibrations, material inhomogeneities, fluctuations in temperature, pressure and refractive index of the air etc influence the cavity stability. The impact of pressure on refractive index fluctuations is largely eliminated by placing the cavity inside a vacuum tank. A properly designed vacuum system also attenuates  acoustic noise. Our focus in this study is on temperature and vibration perturbations which affect the long- and short-term stability of the Fabry-Perot cavity. To achieve a desired tracking accuracy the length changes $\delta L_{1}$ in astro-cavity FP1  must strongly correlate with the length changes $\delta L_{2}$ in lock-cavity FP2. How well  $\delta L_{1}$ and  $\delta L_{2}$ correlate would largely depends on cavity geometry and the way noise is transmitted into the experiment. Normally, the thermal and acoustic noise couples into cavity through mounting and support system. The main purpose of this paper is to estimate noise levels for which the differential length change in two cavities, $\Delta L=\delta L_{1}-\delta L_{2}$ becomes significant. In other words, when $\Delta L\approx0$, the change in RF lock-point of FP2 would accurately reflect the length changes in FP1. Uncorrelated cavity deformations would render the laser-lock technique ineffective.

The shape of an elastic body can deform under the application of a force that may arise from constant acceleration (e.g. gravity loading), floor-vibrations, shocks or acoustic pickup. In most cases, acceleration experienced by a cavity is in the form of vibration perturbations.  These perturbations are time-dependent and contain a large range of frequencies that excite various mechanical eigenmodes of the cavity. The exact nature of interaction depends on  wavelength of excitation relative to the physical size of the cavity \cite{naz06}. In high-frequency limit (small wavelength), a full dynamic analysis is needed to obtain  a complete information of local strain and displacement of the cavity points. Since the excitation wavelengths at higher frequencies are significantly smaller than the cavity size, the conventional numerical discretization methods for dynamic analysis are inefficient and require significant computer resources. Though from practical viewpoint, the high frequency vibrations are less of a problem. They can be effectively attenuated in lab using vibration isolation tables, suitable damping materials, soft cushioning and multi-stage suspension.

In low-frequency limit ($\leq100$ Hz) the dynamic problem of cavity deformation can be reduced to static analysis \cite{che06}. In this regime the excitation wavelength become longer than the cavity size and all points in the cavity respond to the applied force in unison manner. The low-frequency collective motions (compression or stretching) of the particles can be approximated by quasi-static equilibrium resulting from application of a static force or a constant acceleration.  Traditionally,  the vibration or acceleration sensitivity of a cavity is expressed in normalized units of acceleration due to gravity $g$~[ms$^{-2}$]. In this work we also follow the same convention.

\subsection{Mathematical formulism}
Elastic solids change shape under the influence of external force. The elastic deformations can be quantified by knowing the displacement of material points inside the body. Mathematically, elasticity problem can be formulated in terms of field equations which govern the relations among displacements, strains and stresses inside a loaded object. The basic field equations for linear isotropic solids comprise \cite{sad09}:
\begin{itemize}
  \item[] $\bullet$\emph{Strain-displacement relation (compatibility equations)}
      \begin{equation}\label{eq1}
      \varepsilon_{ij}= \frac{1}{2} \left(\frac{\partial U_{i}}{\partial X_{j}}+\frac{\partial U_{j}}{\partial X_{i}}\right)
      \end{equation}
      which completely specifies the strain $\varepsilon_{ij}$ in terms of displacement  $U_{i}(u,v,w)$ and its derivatives. \\

  \item[] $\bullet$\emph{Equations of equilibrium}
     \begin{equation}\label{eq1}
     \frac{\partial \sigma_{ij}}{\partial X_{j}} + F_{i} = 0
      \end{equation}
     which is a statement of Newton's 2nd law for static cases where displacement has no time dependence and derivatives of stress tensor $\sigma_{ij}$  and body forces $F_{i}$ add to zero. \\

  \item[] $\bullet$\emph{Constitutive equations} (Hooke's Law)
     \begin{eqnarray}
     \varepsilon_{ij} &=& \frac{1+\nu}{E} \sigma_{ij}-\frac{\nu}{E}\sigma_{kk}\delta_{ij}
\end{eqnarray}
which establishes general stress-strain behaviour via material parameters, i.e., Young's modulus  $E$ and Poisson's ratio $\nu$.
\end{itemize}

\begin{figure}[htbp]
\centerline{\includegraphics[width=0.75\columnwidth]{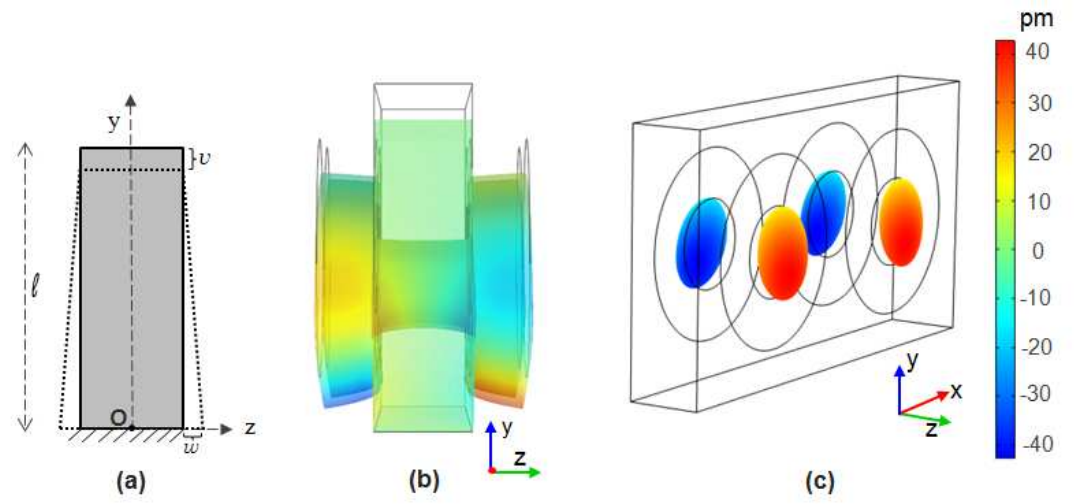}}
\caption{ (a) Compression of a prismatic bar by self-weight. Dotted lines indicate the shape of the bar after deformation. (b) Side view of FP cavity deformed under self-weight. For both cases, the body force is $F_{x}=F_{z}=0, F_{y}=-\rho g$.  The entire structure is vertically compressed and horizontally expanded from original configuration.  (c) The z-displacement and tilt of the cavity mirrors (magnified by $\sim5\times10^{6}$) from their nominal positions. The black outline indicates undeformed shape of the cavity. The color scale represents the cavity displacement in $z$-direction.}
\end{figure}

But for simple geometries, the solution of general system of Eqs.(1)-(3) by analytical methods is almost impossible. Sometimes further simplifications are made by developing a reduced set of field equations solely in terms of stresses or displacements.  One such formulism is to eliminate stresses and strains from Eqs.(1)-(3) and express the displacement in a simplified form, known as Navier's equation. The vector form of Navier's equation is \cite{sad09}:
\begin{equation}\label{eq4}
  \mu \nabla^2 \textbf{U} + (\lambda+\mu)\nabla(\nabla.\textbf{U})+\textbf{F}=0
\end{equation}
where $\mu$ is  shear modulus and $\lambda$ is \textit{Lam\'{e}'}s constant. Equation (4) is usually solved with displacement boundary conditions of the form:  $U_{i}(x_i)=d_{i}(x_i)$, where $x_{i}$ denotes the boundary points on the surface of the body and $d_{i}$ are prescribed displacement values.

As a rough approximation, the cavity deformation problem is similar to  stretching or compression of a prismatic bar when acted upon by an external force. For illustration, we consider a simple rectangular block of length $l$ resting horizontally as shown in Fig. 3(a).  Such a block will be compressed by its own weight with body forces $F_{x}=F_{z}=0$ and  $F_{y}=-\rho g$, where $\rho$ is the mass density and $g$ is acceleration due to gravity. This problem can be solved by direct integration of field equations.  To prevent the translation of the bar we fix the central point $O$ at the lower end of the bar, that is, $x = y = z = 0$ and $u = v = w = 0$. Further, the rotation of the bar about $O$ and about an axis parallel to $O$ is prevented by setting $\partial u/\partial y = \partial w/\partial y = \partial w/\partial x =0 $ at point $O$. Final expressions for three displacement components can be written as:
\begin{eqnarray}
  u &=& \nu \frac{\rho g\,x}{E} (l-y) \\
  v &=& -\frac{\rho g}{2E}\left[2ly -\nu\left(x^2+z^2\right)-y^{2}\right]\\
  w &=&  \nu \frac{\rho g\,z}{E}(l-y)
\end{eqnarray}
where $\rho$ is the material density, $g$ is the acceleration, $\nu$ is Poisson's ratio, $l$ is dimension of the bar along the direction of the force. From Eqs.(5)-(7) and Fig. 3(a) we note that points on the y-axis are only vertically compressed, i.e., $v = -\rho g\left[2ly -y^{2}\right]/2E$. For other points on the bar, the vertical contraction $v$  is accompanied by a finite stretching $u$ and $w$ in  other two orthogonal directions. The coupling of the displacement field is mediated by non-zero value of Poisson's ratio.

\subsection{FEA Simulations}
The example of a prismatic bar illustrated in previous section has an analytical solution due to simple geometry. The actual structure of the FP is quite complex. It has two large cavities and four mirrors attached to the spacer. The dimensions as well as mass of the mirrors are also comparable to the spacer. Using analytical means to find a closed-form solution for complicated geometry is extremely difficult task.  We, therefore, used finite element analysis (FEA) to solve our 3D problem numerically. The FEA is by far the most widely used and versatile numerical technique for simulating deformable solids.
\begin{figure}[htbp]
\centerline{\includegraphics[width=0.55\columnwidth]{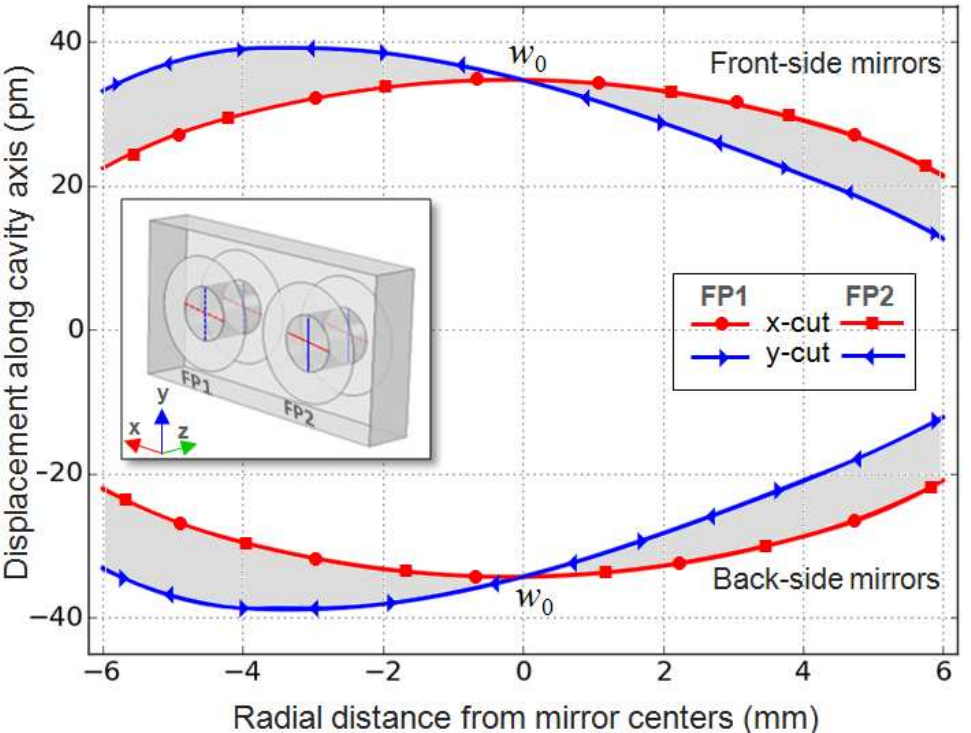}}
\caption{Cavity length changes due to self-weight. In the inset image, horizontal (red) and vertical (blue) lines show the radial probe points defined on the inner surface of each mirror bonded to the spacer block. The red  and blue curves represent the displacement $w$ of horizontal (x-cut) and vertical (y-cut) probe points. The displacement of any other point on  the cavity mirrors is confined to the gray area between the curves. $w_{o}$ denotes the displacement of the mirror centers.}
\end{figure}

We chose COMSOL Multiphysics \footnote{COMSOL Multiphysics Pvt. Ltd., http://www.comsol.com/} as our FEA simulation software. It provides a complete and integrated environment for creating, analyzing and visualizing multiphysics models. Built-in CAD tools and the \emph{Structural Mechanics} module was used for constructing a 3D model of the FP cavity. Additional quantities and model inputs (e.g. boundary constraints, body load, material type and properties etc.) for stress and displacement analysis were defined in \emph{Solid Mechanics} interfaces of the \emph{Structural} module. The relevant material properties used in this study are listed in Table 1. The mesh settings determine the resolution of the mesh element used for
discretizing the model. In all simulations we used default tetrahedral mesh shape. Extra-fine mesh size was chosen to minimize the discretization error. For our study, we selected the stationary solver since our load, stress and deformation do not vary with time.  The \emph{Solid Mechanics} kernel numerically  solves the Navier's equations (Eqs.(4)) and computes displacements, stresses, and strains over the entire FP domain. The deformation of a material point ($x,y,z$) in FP geometry is quantified by calculating the displacement ($u,v,w$) from the undistorted shape. In subsequence analysis, our prime interest will be in displacement component $U_{z}=w$ which is along the axis of the cavity, i.e. $z$-direction.
\begin{table}[htpb]\small
 \centering\caption{Mechanical and thermal properties of the materials used in simulations}\label{tab1}
 \begin{tabular}{l|l l l |l}
         \hline
         Property                       &  \multicolumn{3}{c|}{Material}   & Units \\
                                       &  ULE    & Silica& Zerodur   &   \\ \hline
         Density ($\rho$)              &  2.21   &2.20   & 2.53      & $10^{3}\cdot$kgm$^{-3}$ \\
         Young's modulus ($E$)         &  67.6   &73.1   & 90.3      & $10^{9}$ Pa \\
         Poisson's ratio ($\nu$)       &  0.17   &0.17   & 0.24      & --\\
         CTE ($\alpha$)                &  0.03   &0.55   &-0.08      &  $10^{-6}\textrm{K}^{-1}$ \\
         Thermal conductivity ($k$)    &  1.13   &1.38   & 1.46      &  $\textrm{WK}^{-1} \textrm{m}^{-1}$ \\
         Heat capacity ($C_{p}$)       &  767    &703    & 821       &  $\textrm{Jkg}^{-1}\textrm{K}^{-1}$ \\ \hline
  \end{tabular}
\end{table}
\section{Horizontal Mounting}
The FP is mounted horizontally in $xz$-plane as shown in Fig. 1. The cavity axis is along $z$-direction. Since the cavity length is short we have not considered any discrete support for horizontal mounting. Instead we let the spacer block rest on the bottom plane.

\subsection{Deformation due to uniform gravity load} First we investigate a simple case where the FP unit is deformed under its own weight, i.e., force of gravity $g$ acting uniformly on entire geometry in $-y$ direction.  Fixed boundary constrain $U=0$ is imposed on the bottom $xz$-plane of the spacer to prevent rigid body motion in FEA simulations. Cavity deformation in $yz$-plane due to gravity load is illustrated in Fig. 3(b). In accordance with Eqs. (3)-(5), the gravity-induced compression of the cavity in vertical direction also causes it to expand in $z$-direction, thus increasing the mirror separation. In order to numerically compute the cavity length changes, we extract the displacement component $w$ for each mirror surface that is bonded to the spacer block.  The numerical probe points are defined along $x$ and $y$-direction (see the Fig. 4 inset) on the interior surface of the mirrors forming cavities. The numerical displacement $w$ of the probe points, namely the $x$-cut and $y$-cut, is shown in Fig. 4. In simulations the origin of the coordinate system was chosen to coincides with the geometrical center of the FP spacer block. A near overlap between the FP1 and FP2 probe points indicate that both cavities are deformed identically under the force of gravity.

It is also evident from Fig. 4 that displacement $w$ is not uniform across the cavity cross-section. The horizonal bulge shown by concave shaped red curves, is symmetric about the mirror centers where the displacement is maximum ($\pm36\times10^{-12}$~m)  and reduces to about $\pm24\times10^{-12}$~m towards to cavity edge. The displacement $w$ of the vertical probes, i.e., the $y$-cut shown in blue, is asymmetric. It is minimum ($\pm14\times10^{-12}$~m) at the cavity top edge, attains maxima ($\pm40\times10^{-12}$~m) at about 3~mm below the mirror centers and reduces slightly towards the edges. This asymmetry in displacement is because of the unequal weight along  vertical direction and a small $y$-tilt caused by outward buckling of the spacer and the mirrors. The $z$-displacement of any other point on the mirror surface is bound by gray area between the blue and red curve.

Since the displacement $w$ is not uniform across the reflecting face of the mirror we express the effective cavity length change $\delta L_{1,2}$  for each cavity as the difference in displacement of the mirror centers.  Finally, the quantity of interest for us is differential length change between two FP cavities, i.e. $\Delta L= \delta L_{2}-\delta L_{1}$. For the present case, we find  $\delta L_{1}\approx\delta L_{2}= 75$pm and $\Delta L\approx0$. This is expected as the applied gravity load is symmetric and both cavities are deformed equally.
\begin{figure}[htbp]
\centerline{\includegraphics[width=0.8\columnwidth]{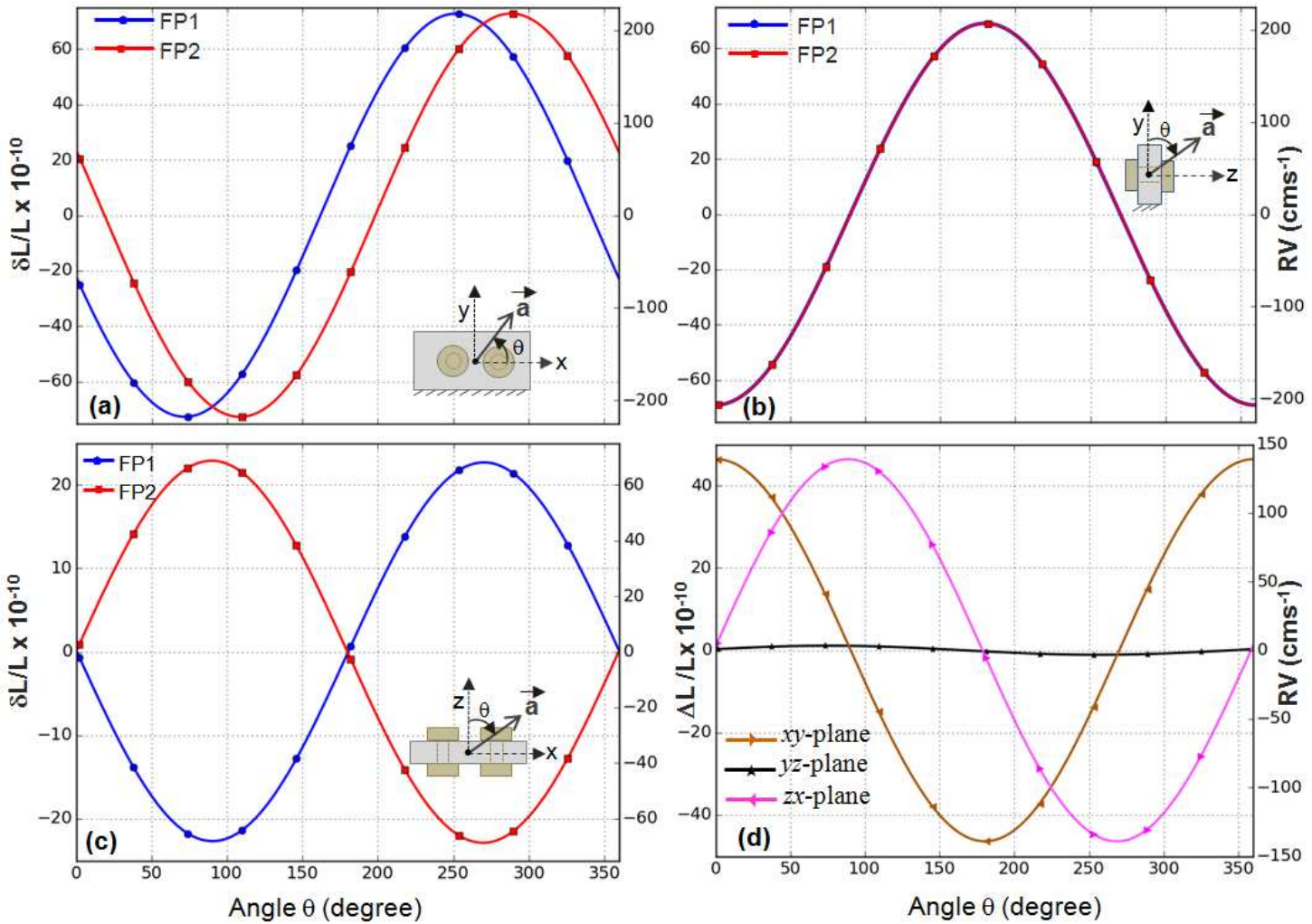}}
\caption{The cavity length change $\delta L_{1,2}$ with gravity-like acceleration vector $\vec{a}$ oriented in (a) $xy$-plane, (b) $yz$-plane and (c) $zx$-plane of the dual cavity. For each plane the angle $\theta$ and corresponding 2D projection of the FP is indicated in the inset figures. The orientation angle $\theta$ was swept from $0-360^{\circ}$ in step of $\Delta\theta=1^{\circ}$.  (d) The differential length change $\Delta L=\delta L_{2}-\delta L_{1}$ between the two FP cavities as a function of sweeping angle. The y-scale to the right gives corresponding RV tracking error (cms$^{-1}$) in each case.}
\end{figure}
\subsection{Orientation Dependent Noise in Horizontal Mounting}
Now we examine cavity's response to a gravity-like static force that is orientation dependent.  This exercise is aimed at understanding  the relative sensitivity of the FP to direction dependent noise perturbations. In simulations, we swept the acceleration vector $\vec{a}$ in different planes and computed differential length change $\Delta L$ for each orientation. The acceleration $\vec{a}$ oriented in a given $ij$-plane, can be expressed in its components along $\hat{i}$ and $\hat{j}$ directions, i.e., $\vec{a}=|\vec{a}|\left(\cos \theta\,\hat{i} + \sin \theta\,\hat{j} \right)$, where for consistency we have chosen $|\vec{a}|=|\vec{g}|$. Using parameter sweep feature of the COMSOL we simulated the impact of different orientation of $\vec{a}$ in the $ij$-plane. Cavity length change  $\delta L_{1,2}$  and $\Delta L$ as a function of orientation in different planes is shown in Fig. 5. The impact of orientation dependent force can be understood by analyzing its response on cavity length separately along each direction.  A few noteworthy observations are:
\begin{itemize}
  \item When the acceleration is along the $x$-direction ($\theta=0^\circ$[$180^\circ$] in Fig. 5(a) and $\theta=90^\circ$[$270^\circ$] in Fig. 5(c)) the sheared strain in FP is produced in such a way that FP1 cavity mirrors are deflected inward [outward] and FP2 cavity mirrors are deflected outward [inward]. Such response can be attributed to unequal shear strain produced because of the constrained boundary and geometrical asymmetry of the cavity in $x$-direction.  A significant differential length change ($\Delta L\approx45$pm) occurs  due to shortening of one cavity  and elongation of the other. This amounts to spurious RV tracking errors about 140 cms$^{-1}$$g^{-1}$ or equivalently a frequency shift of 1.8 MHz$g^{-1}$ at 782 nm laser.

  \item Horizontally mounted dual FP is largely immune to relative length change in vertical vibrations. When the acceleration is oriented in $y$-direction ($\theta=90^\circ$[$270^\circ$] in Fig. 5(a) and $\theta=0^\circ$[$180^\circ$] in Fig. 5(b)) it imparts equal length changes ($\delta L_{1,2}=\pm75$~pm) to each cavity. Since both cavities respond in same way, the relative length change between two cavities is nearly zero. Therefore, a tracking mechanism would measure a correct drift caused by vertical acceleration.

  \item When the acceleration is in $z$-direction ($\theta=90^\circ$[$270^\circ$] in Fig. 5(b) and $\theta=0^\circ$[$180^\circ$] in Fig. 5(c)) each mirror is displaced along the cavity axis by same amount without causing any relative drift i.e., $\Delta L =0$.
  \end{itemize}

\begin{figure}[hbtp]
\centerline{\includegraphics[width=0.5\columnwidth]{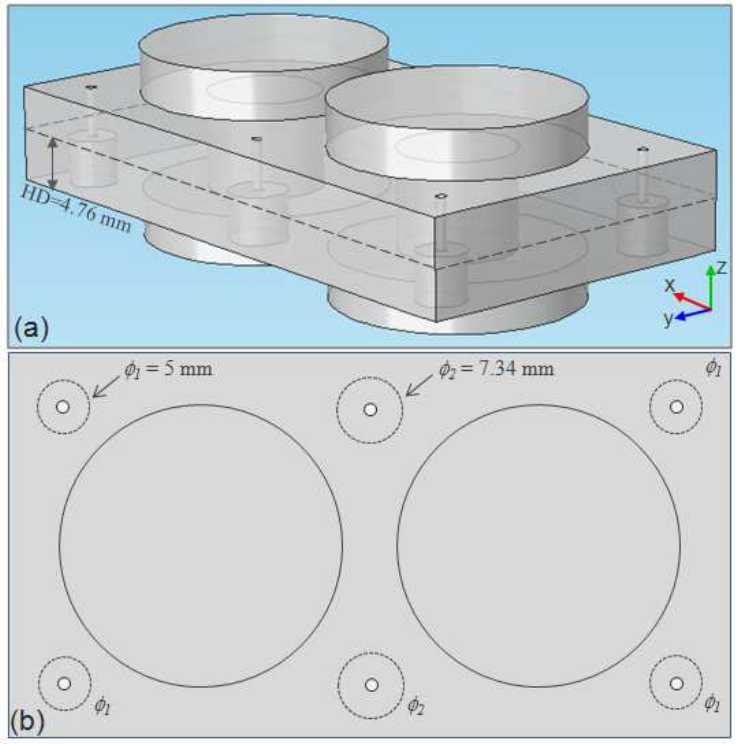}}
v\caption{Vertically mounted dual cavity FP. (a) A 3D view of the geometry with partially drilled mounting holes to provide discrete support-points.  A dashed-line around the spacer block represents the cavity mid-plane. (b) A 2D top-view showing the location of three pairs of mounting holes on the spacer block. The diameter of the corner mounting hole-pairs is denoted by $\phi_{1}$ and the middle mounting hole-pair is denoted by $\phi_{2}$. The concentric through-holes (dia $\approx$ 1 mm) are shown to provide clearance for the vertical suspension with wires. }
\end{figure}

From the proceeding analysis we conclude that proposed FP device is immune to differential length changes if the acceleration noise is oriented along the cavity length and in the vertical direction. However, the noise coupling in $x$-direction can induced non-zero differential length changes ($\Delta L \ne 0$) between two cavities. In such case, the instantaneous drift measured by laser-lock setup will not be same as the drift of astro-cavity. Based on the type of noise some strategies can be adopted to overcome this limitation.  For example, a typical noise spikes (an impulse excitation) only lasts for a fraction of second and temporarily deforms the elastic cavity to produce uncorrelated tracking error in drift measurement. After the noise pulse has elapsed the elastic cavity would return to its original shape and laser-lock would resume the correct measurements. In such case, the actual drift has to be estimated from a series of measurements by carefully discarding the erroneous data. If the source of noise is constant then setup has to be properly shielded. For example, the measurement error due to differential length changes in the proposed device can be reduced below cms$^{-1}$ level by attenuating the noise by $10^{-3}g$.

\section{Vertical Mounting}
The vibration sensitivity of a cavity can be greatly reduced by mounting it vertically. In this scheme the cavity axis is aligned along gravity and FP is  held  close to the spacer mid-plane as illustrated in Fig. 6. The major advantage of this configuration comes from  symmetric distribution of mass above and below the cavity mid-plane. The top and the bottom mirrors are equally displaced by force of gravity in downward direction thus resulting in near-zero change in cavity length. A mounting mechanism can be devised to hold the cavity at suitable supporting positions -also called Airy-points. The cavity spacing can be made invariant by optimizing the size, location and depth of the supporting points.

\begin{figure}[hbtp]
\centerline{\includegraphics[width=0.75\columnwidth]{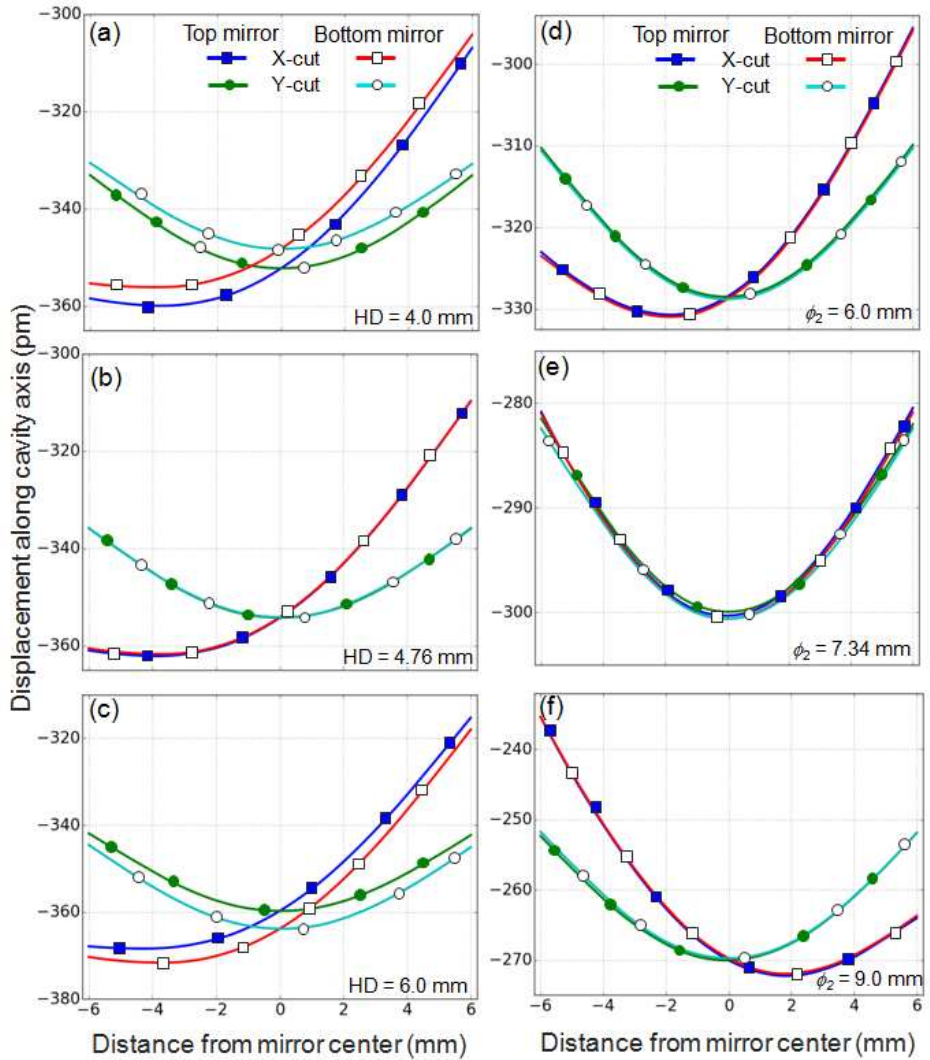}}
\caption{Displacement of top and bottom cavity mirrors in the vertical mounting scheme. (a)-(c) Mirror displacement at different hole depths (HD). In (a)-(c) the diameter of all the mounting holes was fixed i.e. $\phi_{1}=\phi_{2}= 5$ mm and the optimized hole depth was found to be HD=4.76 mm. (d)-(f) Mirror displacement at different diameter $\phi_{2}$ of the central hole pair. In these simulations $\phi_{1}=5$ mm and HD=4.76 mm and  the optimized diameter $\phi_{2}$ of the central hole pair   was found to be 7.34~mm.}
\end{figure}

Three pairs of mounting holes are partially drilled to provide discrete support-points at the under-side of the rectangular spacer. These holes are drilled up to the cavity mid-plane as shown in Fig. 6(a). The flat bottomed-surface of each hole provides a smooth contact area for cavity to rest on mounting posts from below. Alternatively, the FP can be hanged in air by using suspension wires from the above.  The retaining disks attached  to each end of the wire and smoothly  inserted into the mounting holes will wear the cavity weight uniformly.  The concentric through-holes (dia $\approx$ 1 mm) are only shown to provide necessary clearance for wires.  The diameter of the mounting holes at the corners ($\phi_{1}$, see Fig. 6(b)) is chosen to be 5 mm to maintain a sufficient wall thickness from edges.

\subsection*{Optimization of Airy-points}
In vertical mounting Airy-points  are not located exactly at the geometric mid-plane of the cavity. This is because the compression in upper-half of the cavity is not  exactly countered by stretching in lower-half. In addition there is a weight imbalance around the supporting points. The  material removed to form mounting holes reduces the weight of the lower-half of the spacer block.  In order to correctly locate vertical position of the Airy points,  we first varied the mounting hole depth (HD) in our simulations and determined the gravity induced displacement of top and bottom mirrors. The cavity length change is inferred from the displacement of probe points located on the inner surfaces of mirrors as discussed in previous Section.
 \begin{figure}[htbp]
\centerline{\includegraphics[width=0.65\columnwidth]{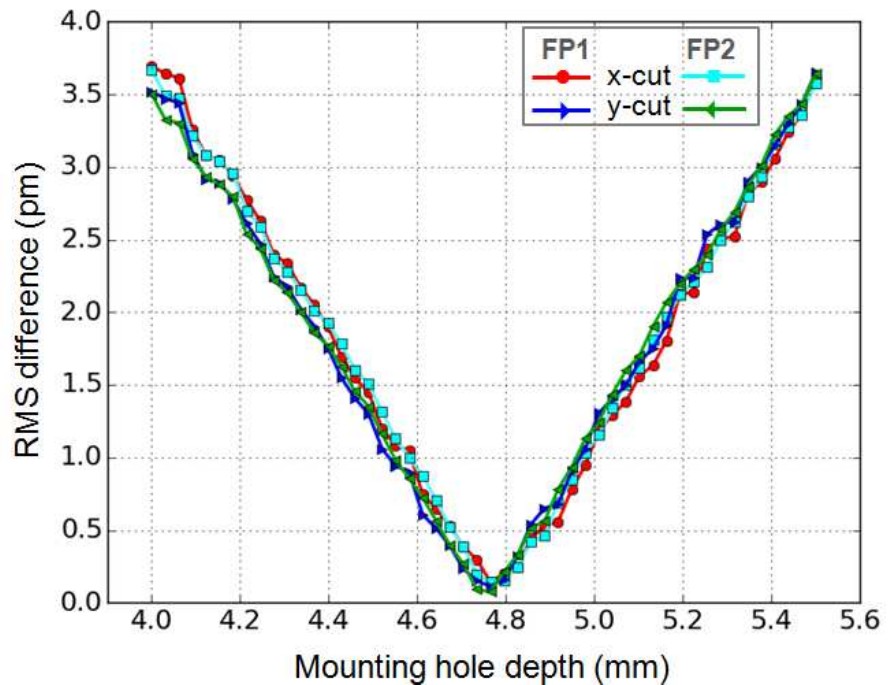}}
\caption{Numerical optimization of the hole-depth (HD) to minimize the gravity induced cavity length changes in vertical mounting. The rms difference in the displacement of top and bottom mirror is less than $ 1.5 \times 10^{-13}$ which corresponds to fractional length change $\delta L/L$  below  $1.5 \times 10^{-11}$.}
\end{figure}

Three test cases with different displacement trends are shown in Fig. 7(a)-(c). When the supporting surface is below the cavity-mid plane, e.g. HD = 4 mm in Fig. 7(a),  the vertical displacement of the upper mirrors is higher as seen from corresponding $x$- and $y$-cuts. The situation is reversed in Fig. 7(c)  where the supporting location  HD = 6 mm  was moved above the  cavity-mid plane. The  null point occurred at HD = 4.76 mm in Fig. 7(b) for which the displacement of top  and bottom mirrors become nearly equal and the cavity length becomes invariant.  Figure 8 shows the rms separation between the top and  the bottom cavity mirrors at various hole depths. The fractional length change at optimal hole-depth for both cavities is below $1.5\times 10^{-11}$. If we assume manufacturing tolerance of  $50\;\mu$m, the rms difference still remains below $10^{-12}$ pm, indicating the robustness of the design against machining errors.  This tolerance limit is well within the current technology of precision manufacturing.

\begin{figure}[htbp]
\centerline{\includegraphics[width=0.9\columnwidth]{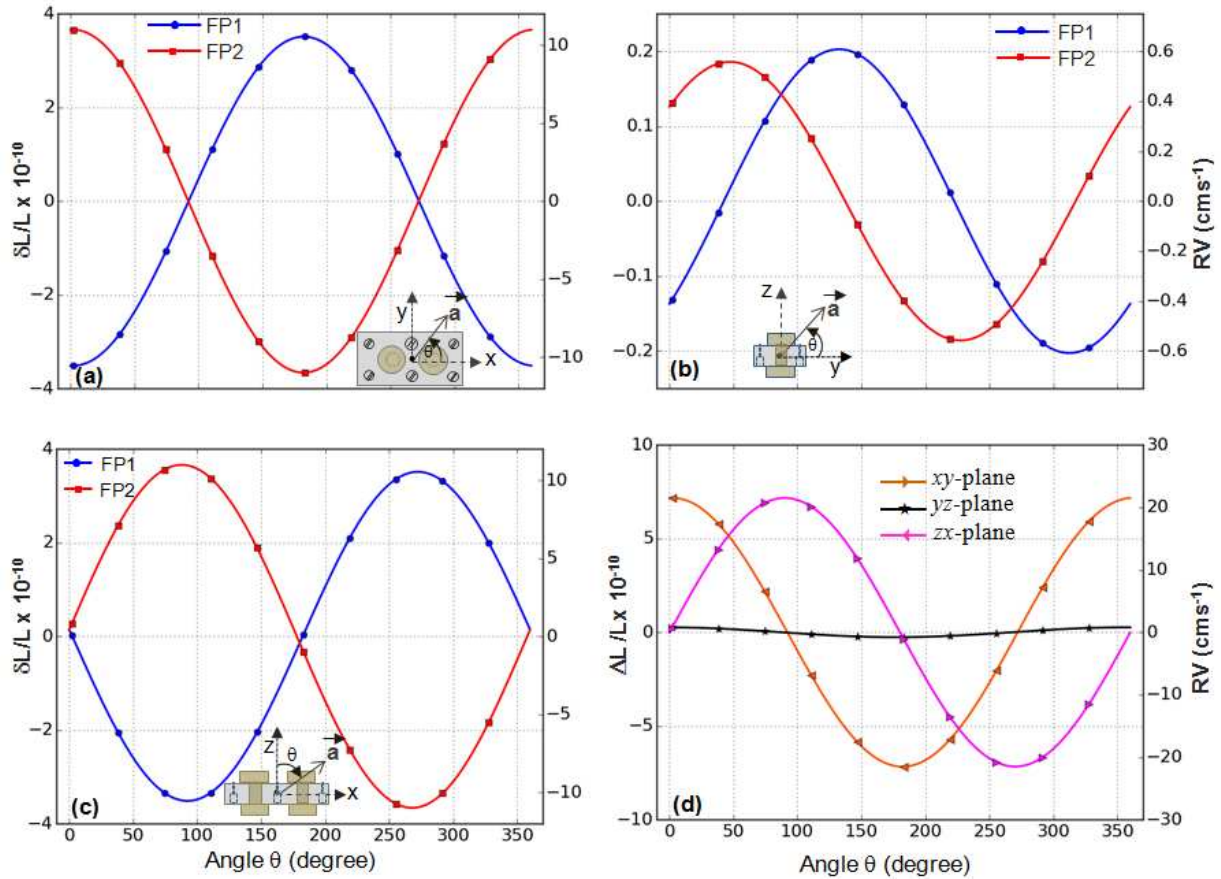}}
\caption{Cavity length variations  $\delta L_{1,2}$ in vertical mounting scheme with acceleration in (a) $xy$-plane, (b) $yz$-plane and (c) $zx$-plane. Coordinate axis are visually depicted in the inset images. For each case the orientation angle $\theta$ was swept from $0-360^{\circ}$ in step of $\Delta\theta=1^{\circ}$.  (d) The differential length change $\Delta L=\delta L_{2}-\delta L_{1}$ between  two FP cavities as a function of orientation  angle $\theta$. The $y$-scale to the right gives corresponding radial velocity error (cms$^{-1}$).}
\end{figure}
Shape of the profile along $x$- and $y$-directions in Fig. 7(a)-(c) also indicates that displacement is not uniform across the cavity cross-section. A linear $x$-tilt is added onto a radially symmetric sag.  This tilt is caused by bending of spacer in the middle that bears larger mass concentration. We enlarged the diameter $\phi_{2}$ to remove extra material for balancing and equalizing the weight distribution in middle part of the spacer. Figure 7(e)-(f) demonstrates how the tilt component (x-cut) changes sign  when $\phi_{2}$ was varied from 6 mm to 9 mm. At optimum diameter ($\phi_{2} = 7.34$ mm in Fig. 7 (e)), the tilt component was largely eliminated and the rms difference between $x$ and $y$-radial profile was reduced below $1\times10^{-12}$ m.

Having found an equilibrium plane for support, we also estimated the direction dependent noise sensitivity of the vertical mount. For FEA simulations, the mechanical  displacement of the mounting holes was constrained and the FP geometry was subjected to a sweeping acceleration vector in different planes.  The calculated cavity length variations $\delta L_{1,2}$ and differential length change $\Delta L$ is plotted in Fig. 9. For acceleration along $x$, the vibration sensitivity to differential length change has reduced by a factor of seven compared to horizontally mounted case. For acceleration along $y$ and $z$-direction, the RV error is below 1 cms$^{-1}$g$^{-1}$. Since the vibration sensitivity is significantly reduced in vertical mounting scheme, the requirements for external noise isolation and damping are also relaxed.

\section{Temperature Sensitivity}
Thermally induced material expansion or contraction is also a major source of noise against the dimensional stability of optical materials. A desired length stability of FP cavity can be assured by choosing  appropriate spacer/mirror material with low thermal expansion coefficients. To minimize the impact of temperature fluctuations, the reference cavities are also operated near their zero-crossing temperature \cite{leg10}. The entire setup is enclosed inside an evacuated and thermally shielded environment. In this section  we investigate effect of temperature fluctuations on our FP cavities. To  first approximation, the cavity length changes in FP1 and FP2 will strongly correlate if temperature variations within the enclosure are uniform. In that case the tracking mechanism would expectedly register a correct drift.  However, small temperature gradients inside the enclosure will expose two cavities to a locally different environment, producing dissimilar length changes in FP1 and FP2.  This would contaminate the tracking signal with intractable radial velocity errors.

\begin{figure}[htbp]
\centerline{\includegraphics[width=0.9\columnwidth]{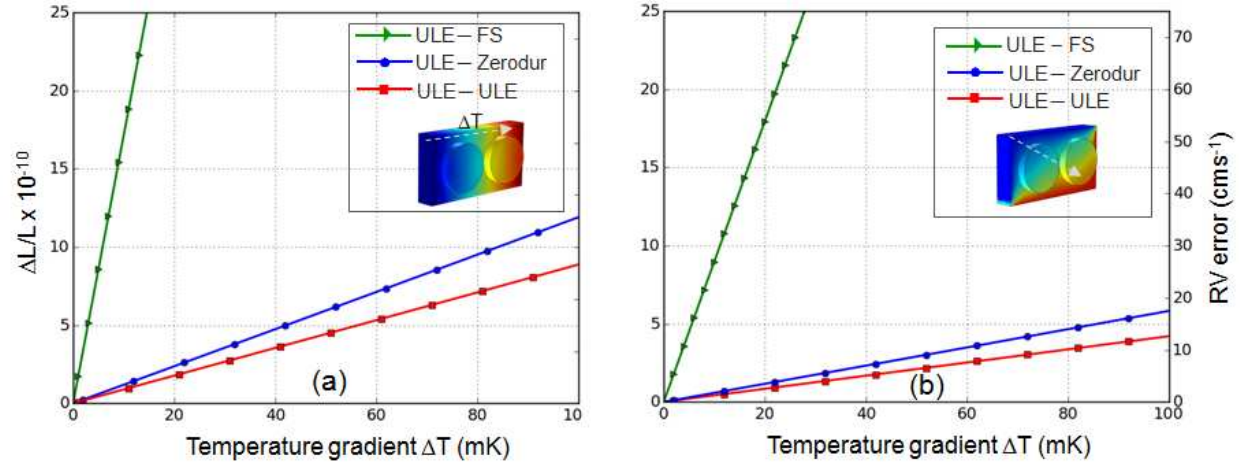}}
\caption{The differential length change due to temperature gradient (a) along the spacer-edge and (b) along the diagonal direction. The arrow  shows the direction  of gradient. Corning ULE was used as a spacer material with Fused Silica (FS), Zerodur and ULE as mirror substrate. The magnitude of the temperature gradient $\Delta T$ was varied from 0 mK to 100 mK.}
\end{figure}
\subsection{Effect of Temperature Gradients}
To analyze the temperature sensitivity of the FP we again used FEA to solved the thermal-stress problem with COMSOL multiphysics. In our thermal model, realistic temperature gradients were simulated by imposing appropriate temperature conditions on selected boundaries of the spacer block. A steady-state numerical solution then establishes a linearly varying temperature  and stress fields between the selected sides. Temperature sensitivity also depends on the choice of the spacer and mirror material. In these studies, Corning ULE was used as cavity spacer while three glass-ceramics namely, Zerodur, Silica and Corning ULE were tested for mirror substrates. Two worse case scenario  producing unequal cavity length changes are shown in Fig.~10. The temperature sensitivities ($\Delta L /L$ per mK) of the device can be determined from the slopes of the plots. Among the chosen  spacer-mirror pairs, the ULE-FS cavity has the largest sensitivity $1.71\times10^{-10}$~mK~$^{-1}$ (RV error $\approx$ 3.81~cms~$^{-1}$~mK~$^{-1}$) while ULE-ULE has the least sensitivity $0.09\times10^{-10}$~mK~$^{-1}$ (RV error $\approx$  0.27 cms$^{-1}$mK$^{-1}$) for temperature gradient shown along spacer edge in Fig. 10(a). As see in Fig. 10(b), the temperature sensitivity is almost halved  for the gradient considered along diagonal direction.

The contribution of CTE to cavity length change from mirror substrate is substantially large compared to the contribution from spacer material. The main drawback of fused silica is its large CTE which places stringent requirements on the design of thermal control system of the cavity. For example, to keep the RV error of dual cavity FP below 1~cms$^{-1}$ level, a temperature uniformity at sub-mK level has to be ensured. This requirement is relaxed to few-mK  if both spacer-mirror pair are made with extremely low expansion materials such as Zerodur and ULE. A new method of building temperature insensitive FS cavity mirrors is expected improve their performance in precision metrology \cite{leg10}. Despite large CTE, fused silica is considered  as best performing material for cavity mirrors.  This is because FS has very small mechanical losses at room temperature which helps to lower cavity's thermal noise floor which is about 2-3 times better compared to the ULE or Zerodur \cite{num04}.

\subsection{Effect of Temperature on Unequal Cavity Spacing}
 In ideal case of equal length cavities i.e $L_1=L_2$, temperature fluctuations will not produce any relative length change between two cavities. In practice, it is not possible to manufacture cavities of equal length with arbitrary precision. A small offset in lengths may arise from  machining errors and/or imperfections in the optical bonding  between the mirrors and the spacer. To study this effect we created cavities with different length offset and calculated their differential expansion from the thermal model. The length offset was introduced by shortening the spacer thickness $L_1$ of FP1 without altering $L_2$. In the FEA simulations the temperature of the entire FP block was varied uniformly in step of 1 mK. Figure 11 shows thermally induced differential length changes between two cavities for different length offsets. Clearly, a 0.1 K difference from the temperature set point can produce $\sim3$ pm  differential expansion (the green curve in Fig. 11) between two cavities if the initial length offset was $100\;\mu$m. This amounts to $\sim10$~cms$^{-1}$ RV tracking error. These calculations reinforces the need of high thermal stability (few mK) and cavity length matching better than $50\;\mu$m  for keeping the influence of temperature changes on RV error below 1~cms$^{-1}$.

\begin{figure}[htbp]
\centerline{\includegraphics[width=0.65\columnwidth]{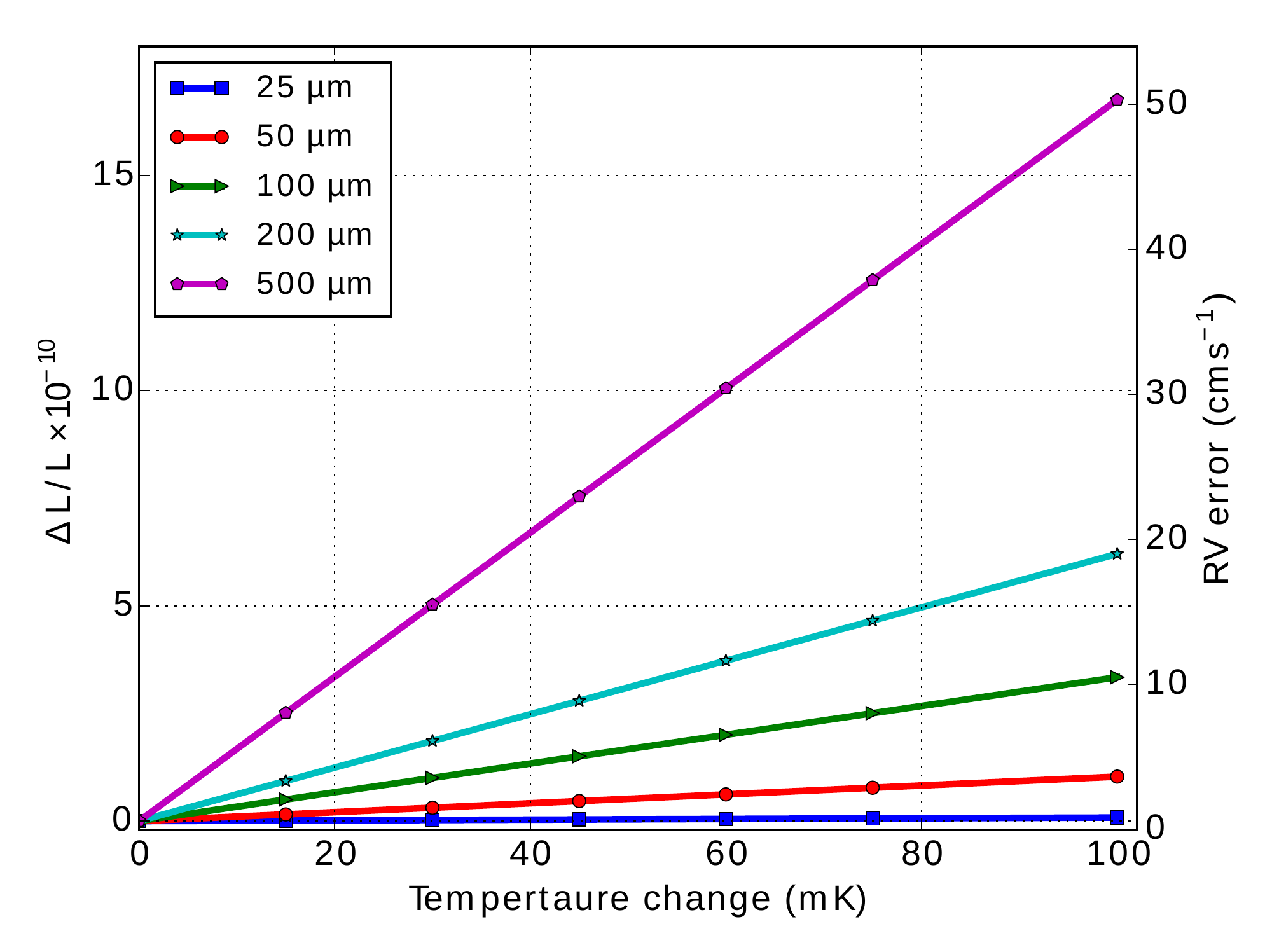}}
\caption{Temperature induced differential length changes for cavities of unequal length. Temperature of the entire FP block was varied uniformly from $T_0=305$ K to $T_{0}$ + 0.1 K in incremental step of 1 mK. The spacer material used in simulation is ULE and mirror material is fused silica. Various length offsets $|L_2-L_1|$ are given in $\mu$m.}
\end{figure}
\section{Numerical Accuracy}
In FEA, the numerical solution of a problem is approximated by partitioning the CAD geometry into smaller parts. The accuracy of the numerical solution is intrinsically linked to the mesh size. By refining the mesh, the solution becomes more accurate but this is achieved at the cost of finite computational resources and time. In a realistic FEA model, the difference between exact and  approximate solution should be minimized and error should not exceed the tolerance limit defined by some accepted criteria.

We used the default tetrahedral mesh settings with adaptive mesh refinement in our FEA software to discretize the cavity model. The refinement algorithm globally adjusts the mesh size  to provide sufficient resolution around all regions of interest within the geometry. There is no generic rule to verify the numerical accuracy of FEA model -especially in complex problems where exact solution is not known a priory. In order to evaluate the discretization error due to finite mesh size, we compare the relative displacement $w$ of FP1 and FP2 mirrors for a simple gravity loading case (see Fig. 4) already discussed in Section 4.1. From the application of uniform load and symmetry consideration,  exact solution requires the cavity displacement in FP1 and FP2 to be equal. Any difference in displacement can be attributed to numerical inaccuracies originating from meshing errors. Figure 12 shows the  displacement data ($x$-cut) plotted for input cavity mirrors at three different mesh densities.

We note that compared to the smaller mesh sizes, the magnitude of the displacement is slightly underestimated at \emph{`coarse mesh'}. In addition, a small relative difference can also be seen in Fig. 12(a).  This difference reduces progressively with decreasing mesh size (see Fig. 12(b)) and becomes insignificant for `\emph{extra-fine mesh}' shown in Fig.12(c) where two curves completely overlap. In each case the numerical error is estimated from the rms difference between the two curves. The discretization rms error for extra-fine mesh is $3.6\times10^{-14}$ m. This amounts to $<4\%$ error for cavity length stability targeted at $10^{-12}$ m level. All FEA simulations  in this paper were carried out with extra-fine mesh.

Another possible source of error is uncertainties in thermal and mechanical properties of the ceramic
material listed in Table 1. Typically, the material parameters can have about $2\%$ deviation from their nominal values. We varied the material parameters in our simulation by $5\%$ but did not find any noticeable difference in vibration or temperature sensitivities. The calculated deviation was of the same order as the discretization errors. This is not unexpected  since our FP is lightweight and cavity length is short. For the same level of uncertainties in Poisson ratio, Youngs modulus and material density, a vibration sensitivity of $\approx 10^{-13}g^{-1}$  was achieved in longer cavities \cite{naz06}.

\begin{figure}[htbp]
\centerline{\includegraphics[width=0.65\columnwidth]{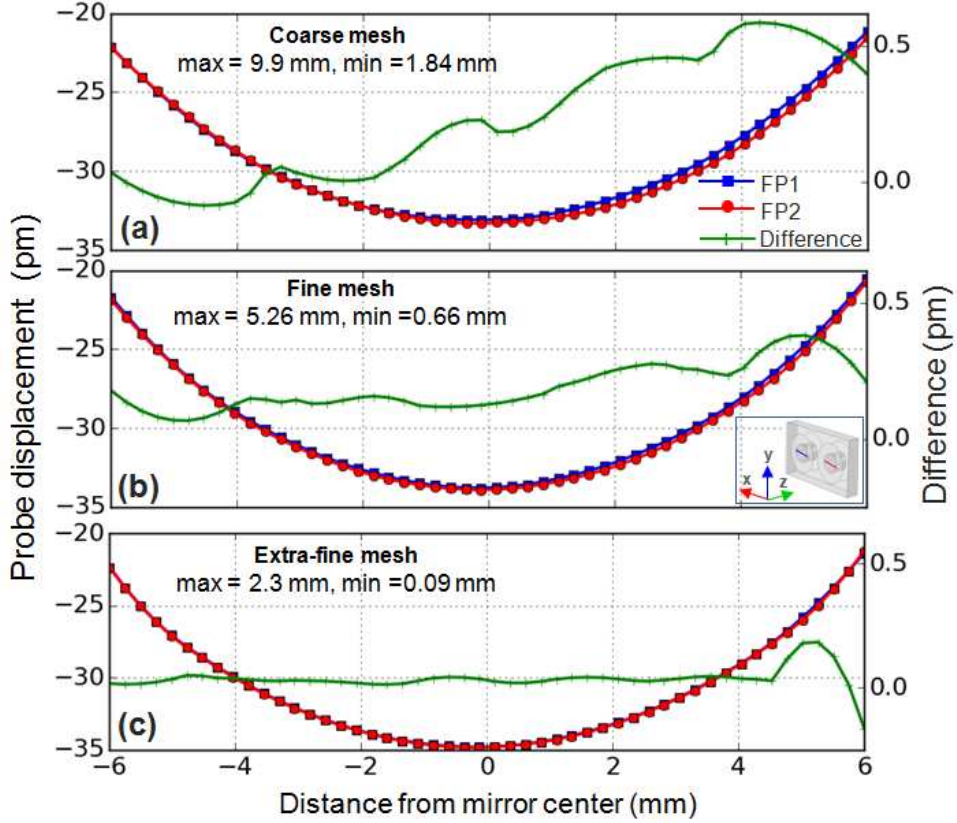}}
\caption{Displacement due to uniform gravity loading obtained at three mesh densities: (a) a coarse mesh (b) a fine mesh and (c) an extra-fine mesh. Inset image shows the $x$-data probes on the input mirrors of FP1 and FP2 for which the cavity displacement is numerically computed. The default element size (min, max) for each case in also indicated. The green curve ($y$-scale to the right) shows the difference between the two plots. The RMS error for coarse, fine and extra-fine meshing is $32.0\times 10^{-14}$ m, $28.0\times 10^{-14}$ m and $3.6\times 10^{-14}$ m, respectively. }
\end{figure}

\section{Summary}
The Doppler survey for many astronomy programmes requires high resolution spectrograph with exceptional long term stability in RV precision.  In this paper we have proposed  a double cavity FP design for aiding high precision Doppler measurements in astronomy. The proposed device has two parallel optical cavities embedded inside a 10~mm thick monolithic ULE block. A low-finesse \emph{astro-cavity} is meant to generate multiline reference spectra for  spectrograph calibration while a high finesse \emph{lock-cavity} is needed to provide steep discriminant signal for tracking the cavity drift. External noise can have severe detrimental effect on the performance and reliability of the tracking measurements. An accurate measurement of drift is possible only if the noise interventions are common to both cavities.

We have used FEA to analyze  FP's sensitivity to mechanical vibrations and temperature fluctuations.  A 3D FEA simulations were carried out for the FP structures comprising five  domains (4 mirrors + 1 spacer). An optimized cavity design with proper mounting and support system helps insulate the noise. Two mounting geometries were investigated for tracking errors resulting from differential length changes between the cavities. The horizontal geometry showed vibrational sensitivity up to $\sim 45\times10^{-10}$$g^{-1}$ (RV errors $\sim135$ cms$^{-1}$) for acceleration directed along the longer edge ($x$-axis) of the block. The noise along $y$- and $z$-direction is common and does not contribute to differential length change. For achieving 1 cms$^{-1}$ RV stability of the device the horizontal mounting configuration may require a well thought strategy to insulate the vibration noise down to $\sim 10^{-3}g$.

Cavity lengths can be made invariant by mounting the FP vertically. We worked out a vibration insensitive design by numerically optimizing the hole-depth and diameter of the mounting holes. In the optimized design the tilt-component was  eliminated and a clear null point was achieved for gravity induced displacement between the mirrors.

We also studied the impact of thermal instability and choice of mirror/spacer material on the tracking performance. A steady-state thermal analysis was carried out to evaluate the effects of thermal gradient. Numerical simulations show that temperature gradient within the FP enclosure has to be accurately controlled and relative length offset between two cavities should be minimized to avoid differential length changes that may cause spurious drift and unintended tracking errors. The proposed dual FP device can be easily fabricated with available machining precision of $50-100\;\mu$m and readily achievable thermal stability of few mK.

This work is also relevant to cavities used in laboratory experiments designed to test Lorentz invariance where a pair of orthogonal FP cavities is employed to detect anisotropy in the speed of light \cite{her09}. The entire optical setup is put on a rotating table. A  frequencies beat-note between two lasers, each locked to one of the two orthogonal cavities, is examined while the entire setup is continuously rotated on the turntable. Any possible anisotropy is expected to modulate the laser beat signal at twice the rotation rate. However, the discrimination of the anisotropic signal from competing noise artifacts is not trivial. Even though the length of two cavities is defined by a single ULE block, detection ambiguities may arise, e.g., from periodic deformation of cavity length caused by gravitational and centrifugal forces. The approach outlined in our studies might be useful to understand and model various systematics producing relative drift  in cross-cavities.

\section*{Acknowledgement} RKB  acknowledges the research support for this project (reference no. EMR/2014/000941) from Science and Engineering Research Board (SERB), Department of Science and Technology (DST), INDIA.  A.R. acknowledges the research support from European Research Council under the FP7 Starting Grant agreement number 279347 and funding from DFG grant RE 1664/9-1.

\end{document}